\documentclass[aps,prb,groupedaddress,showpacs,twocolumn]{revtex4}
\usepackage{graphicx}
\usepackage{amsmath}
\usepackage{bbm}
\usepackage{xspace}
\usepackage{color}

\newcommand{\se}{Sec.\@\xspace}
\newcommand{\Se}{Section\@\xspace}

\newcommand{\ie}{i.\thinspace{}e.\@\xspace}
\newcommand{\nag}{{\phantom{\dag}}}

\newcommand{\ptl}{\partial}

\newcommand{\PDF}[2]{\frac{\ptl\, #1}{\ptl\, #2}}

\newcommand{\ve}[1]{{\bf #1}}
\newcommand{\mat}[1]{\mathsf{#1}}

\newcommand{\eq}[1]{Eq.\thinspace{}(\ref{#1})}

\newcommand{\tab}[1]{Tab.\thinspace{}\ref{#1}}

\newcommand{\fig}[1]{Fig.\thinspace{}\ref{#1}}

\newcommand{\fc}[1]{({#1})}

\newcommand{\Fig}[1]{Figure\thinspace{}\ref{#1}}

\newcommand{\Tr}{\mbox{Tr}}

\def\bra#1{\mathinner{\langle{#1}|}}
\def\ket#1{\mathinner{|{#1}\rangle}}

\begin{document}


\title{Quantum phase transition and excitations of the Tavis-Cummings lattice model}


\author{Michael Knap}
\email[]{michael.knap@tugraz.at}
\affiliation{Institute of Theoretical and Computational Physics, Graz University of Technology, 8010 Graz, Austria}
\author{Enrico Arrigoni}
\affiliation{Institute of Theoretical and Computational Physics, Graz University of Technology, 8010 Graz, Austria}
\author{Wolfgang von der Linden}
\affiliation{Institute of Theoretical and Computational Physics, Graz University of Technology, 8010 Graz, Austria}


\date{\today}

\begin{abstract}
The enormous progress in controlling quantum optical and atomic systems has prompted ideas for new experimental realizations of strongly correlated many-body systems operating with light. These systems consist of photons confined in optical cavities, which interact strongly with atoms or atomiclike structures. Due to the interaction between the two particle species optical nonlinearities appear, leading to a quantum phase transition from Mott to superfluid phase. Here, we address the Tavis-Cummings lattice model, which describes light-matter systems containing multiple atomiclike structures in each cavity. In particular, we investigate the phase boundary delimiting Mott from superfluid phase and the elementary excitations of the two-dimensional Tavis-Cummings lattice model in dependence of the number of atomiclike structures per cavity. In order to obtain the results we employ the variational cluster approach. We evaluate spectral functions and densities of states of both particle species, which allows us to characterize the fundamental excitations of light-matter systems. These excitations are termed polaritons and are superpositions of photons and atomic excitations. We introduce polariton quasiparticles as appropriate linear combinations of both particle species and analyze the weights of their constituents. Our results demonstrate the dependence of the quantum phase transition and the elementary excitations on the number of atomiclike structures per cavity and provide thus valuable insight into the physics of light-matter systems.
\end{abstract}


\maketitle

\section{\label{sec:introduction}Introduction}
The push towards the experimental realization of quantum computers lead to incredible advances in the fields of quantum optics and atomic physics. Unprecedented experimental control in these fields allowed to envision new realizations of strongly correlated many-body systems, which operate with light.\cite{hartmann_strongly_2006, greentree_quantum_2006, angelakis_photon-blockade-induced_2007, hartmann_quantum_2008} Confined light modes in coupled cavity arrays are able to tunnel between adjacent sites and thus propagate on a lattice of cavities. Strong correlations in turn can be observed when a repulsive interaction between photons is present.\cite{strong}  
This repulsion, which is termed optical nonlinearity, can be achieved by coupling the light modes to matter in the form of atoms or atomiclike structures present within each cavity. The interaction between the light modes and atomic like structures is achieved by means of dipole coupling. In theory there exist two major schemes to obtain this interaction. The first is to model the atomiclike structures by two-level systems, leading to an interaction of the Jaynes-Cummings type,\cite{jaynes_comparison_1963, greentree_quantum_2006} 
whereas the second approach is based on electromagnetically induced transparency\cite{boyd_photonics:_2006} 
and uses four-level systems.\cite{hartmann_strongly_2006} In both cases optical nonlinearities between photons arise as the energy for adding two photons to the cavity is larger than twice the energy needed to add one photon. This behavior leads to intriguing experiments such as the photon blockade effect,\cite{birnbaum_photon_2005} where only one elementary excitation is present in the cavity at the same time. The elementary excitations in light matter systems are termed polaritons. Polaritons are superpositions of both particle species, namely photons as well as excitations of the atomiclike structures. Following these considerations and arranging multiple cavities on a lattice leads to a strongly correlated phase in which photons are involved. As a result, light-matter systems exhibit a quantum phase transition from Mott phase where polaritons are localized in the cavities to superfluid phase where polaritons are delocalized on the whole lattice.\cite{hartmann_strongly_2006, greentree_quantum_2006, angelakis_photon-blockade-induced_2007} 

Up to now an experimental realization of light-matter systems is still missing, however, there are several promising approaches such as quantum dots grown in photonic crystal cavities, transmission line cavities and toroidal or disk shaped cavities.\cite{hartmann_quantum_2008} The advantage of light-matter systems is that they are of mesoscopic size and thus allow for a direct addressability of each lattice site and good experimental control on the system parameters. Exhibiting these valuable properties light-matter systems might be used as quantum simulators for other strongly correlated many-body systems such as the Bose-Hubbard model,\cite{fisher_boson_1989} or find their applications in quantum information processing.

Experimentally it might be more feasible to place multiple atomiclike structures within one cavity. Therefore it is important to study light-matter systems which contain more than one two-level system per cavity. The theoretical model describing a single cavity with $N$ two-level systems is termed Tavis-Cummings model.\cite{Tavis_Exact_Solution1968,Tavis_Approximate_Solutions_1969} In the case of $N=1$ it reduces to the Jaynes-Cummings model.\cite{jaynes_comparison_1963} Light-matter systems with coupled Jaynes-Cummings cavities have already been investigated to some detail in Refs.~\onlinecite{greentree_quantum_2006, rossini_mott-insulating_2007, rossini_photon_2008, aichhorn_quantum_2008, makin_quantum_2008,
irish_polaritonic_2008, koch_superfluid--mott-insulator_2009, schmidt_strong_2009, pippan_excitation_2009, knap_jcl_2010} and will thus not be addressed here anymore. However, systems of coupled-cavity arrays with more than one two-level system per cavity have been rarely studied in literature. In particular, the quantum phase transition has been investigated on mean field level by N.~Na \textit{et al.} in Ref.~\onlinecite{na_strongly_2008}, and in one-dimension with density matrix renormalization group (DMRG) by D.~Rossini \textit{et al.} in Ref.~\onlinecite{rossini_mott-insulating_2007}. In the present paper, we investigate the quantum phase transition and the elementary excitations---the polaritons---of coupled Tavis-Cummings cavities arranged on a two-dimensional lattice. In particular, we evaluate the phase boundary delimiting Mott phase from superfluid phase for different number of two-level systems per cavity. Furthermore, we study spectral properties of both photons as well as atomic excitations which in turn allows us to characterize the polaritonic properties of the system. In order to evaluate the quantum phase transition and the spectral excitations we employ the variational cluster approach.\cite{potthoff_variational_2003}

This paper is organized as follows. The Tavis-Cummings lattice model is introduced in \se~\ref{sec:model}. Details on the variational cluster approach are covered in \se~\ref{sec:method}. In \se~\ref{sec:qpt} we discuss the results obtained for the phase boundary delimiting Mott from superfluid phase. \Se~\ref{sec:excitations} is devoted to the excitations of the Tavis-Cummings lattice model. Here, we present spectral functions and densities of states of both particle species and discuss polariton quasiparticle excitations. Finally, we summarize and conclude our findings in \se~\ref{sec:conclusion}.

\section{\label{sec:model}The Tavis-Cummings lattice model}
\begin{figure}
	\centering
	\includegraphics[width=0.48\textwidth]{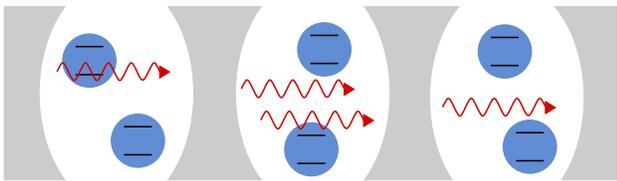}
	\caption{(Color online) Illustration of the TCL model. Blue bubbles represent two-level systems and red wavy arrows photons. Photons and two-level system interact via dipole coupling. }
	\label{fig:tcl}
\end{figure}

A single cavity at lattice site $i$ containing $N$ two-level systems is modeled by the Tavis-Cummings (TC) Hamiltonian,\cite{Tavis_Exact_Solution1968,Tavis_Approximate_Solutions_1969}
\begin{equation}
 \hat{H}^{TC}_i = \omega_c \, a_i^\dagger \, a_i^\nag + \epsilon \, (S_i^z + \frac{N}{2} ) + g \,( a_i^\nag\,S_i^+ + a_i^{\dagger}\,S_i^- ) \;\mbox{,}
 \label{eq:tc}
\end{equation}
where $\omega_c$ is the resonance frequency of the cavity, $\epsilon$ is the energy spacing of the two-level systems, and $g$ is the atom-field coupling constant (see a single cavity in \fig{fig:tcl} for illustration). 
The operators $a_i^\dagger$ and $a_i^\nag$, respectively, create and annihilate photons in the cavity $i$. The ensemble of two level systems is described by collective spin operators $S_j^\alpha = \sum_{\nu=1}^N \sigma_{\nu,j}^\alpha$, where $\alpha \in \lbrace z,\,+,\,- \rbrace$, and $\sigma_{\nu,j}^\pm = \sigma_{\nu,j}^x \pm i \sigma_{\nu,j}^y $ are the spin raising and lowering operators. When starting from the dipole interaction between photons and two-level systems two additional terms arise in the Hamiltonian, which are proportional to $a_i^\dagger\,S_i^+$ and $a_i^\nag\,S_i^-$. However, for the condition $|\omega_c-\epsilon| \ll \omega_c,\,\epsilon $ these terms are fast oscillating in comparison to $a_i^\nag\,S_i^+$ and $a_i^{\dagger}\,S_i^-$ and can thus be neglected, which is known as rotating wave approximation.\cite{haroche_exploringquantum:_2006}
The difference between the resonance frequency of the cavity $\omega_c$ and the energy spacing of the two-level system $\epsilon$ is termed detuning $\Delta=\omega_c-\epsilon$. As a consequence of the rotating wave approximation the total number of excitations $\hat{n}_i = a_i^\dagger \, a_i^\nag + S_i^z + N/2$ is conserved. Additionally, the total spin $S^2$ is a conserved quantity as well. The ground state of the TC model is always in the sector of maximum spin $S=N/2$,\cite{rossini_mott-insulating_2007} which will thus be considered in further calculations.

The full model consists of $M$ coupled cavities, which are arranged on a lattice. Therefore we refer to this model as Tavis-Cummings lattice (TCL) model. The TCL Hamiltonian is given by 
\begin{equation}
 \hat{H}^{TCL} = -t \sum_{\left\langle i,\,j \right\rangle} a_i^\dagger \, a_j^\nag + \sum_i \hat{H}_i^{TC} - \mu\,\hat{N}_p \;\mbox{,}
 \label{eq:tcl}
\end{equation}
where the first term allows photons to tunnel between cavities $i$ and $j$. The tunneling strength $t$ is given by the overlap integral of the photonic wave functions, which is considered to be nonzero only for nearest-neighbor sites $i$ and $j$. The restriction to nearest neighbors is denoted by the angle brackets $\langle \cdots \rangle$ around the summation indices. The second term of \eq{eq:tcl} describes the physics of the individual cavities and the last term controls the average particle number of the system, where $\mu$ is the chemical potential and $\hat{N}_p=\sum_i \hat{n}_i$ is the total particle number. \Fig{fig:tcl} illustrates the TCL model.  For the TCL model the total particle number $\hat{N}_p$ is conserved as well as the total spin $S^2$ of each cavity. As in the case of the Jaynes-Cummings lattice model\cite{jaynes_comparison_1963} 
the TCL model can be rewritten as
\begin{align}
 \hat{H}^{TCL} = &-t \sum_{\left\langle i,\,j \right\rangle} a_i^\dagger \, a_j^\nag -\Delta \sum_i (S_i^z + \frac{N}{2})  \nonumber \\
&+g\sum_i  ( a_i^\nag\,S_i^+  + a_i^\dagger\,S_i^- ) - (\mu-\omega_c)\,\hat{N}_p\;\mbox{.}
 \label{eq:spe:tclrewritten}
\end{align}
In the forthcoming discussions and calculations we use the dipole coupling $g$ as unit of energy. Therefore the physics of the TCL model depends only on three independent parameters, namely, the hopping strength $t$, the detuning $\Delta$, and the modified chemical potential $\mu-\omega_c$.

\section{\label{sec:method}The variational cluster approach}
In order to evaluate the boundary of the quantum phase transition from Mott phase to superfluid phase and to investigate the excitations of the TCL model we employ the variational cluster approach\cite{potthoff_variational_2003, potthoff_self-energy-functional_2003-1, potthoff_self-energy-functional_2003, koller_variational_2006} 
(VCA), which yields the single-particle Green's function $G(\ve k, \,\omega)$ of the physical system $\hat{H}^{TCL}$. VCA has been previously applied to light-matter systems in Refs.~\onlinecite{aichhorn_quantum_2008, knap_jcl_2010}.

The basic idea of VCA is that the grand potential $\Omega$ can be expressed as a functional of the self-energy $\mat \Sigma$ and that Dyson's equation for the Green's function is fulfilled at the stationary point of $\Omega[\mat \Sigma]$. To be able to evaluate $\Omega[\mat \Sigma]$, the unknown self-energy $\mat \Sigma$ of the physical system is approximated by the self-energy of an exactly solvable, so-called reference, system. In VCA the reference system is chosen to be a cluster decomposition of the physical system, which means that the system of size $M$ is divided into clusters of size $L$ with, however, different single-particle parameters $\mat{x}$ as compared to the physical system. Due to the approximation in the self-energy the functional $\Omega[\mat \Sigma]$ becomes a function of the single-particle parameters $\mat x$ of the reference system
\begin{equation}
 \Omega(\mat{x}) = \Omega^\prime(\mat{x}) + \Tr\,\ln(-\mat{G}^{\prime}(\mat{x})) + \Tr\,\ln(-(\mat{G}_0^{-1}-\mat{\Sigma}(\mat{x}))) \;\mbox{,}
\label{eqn:om}
\end{equation}
where quantities with prime correspond to the reference system and $\mat{G}_0$ is the noninteracting Green's function. The stationary condition on $\Omega (\mat{x})$ now reads
\begin{equation}
 \PDF{\Omega (\mat{x})}{\mat{x}} = 0 \; \mbox{,}
 \label{eq:stat}
\end{equation}
which can be evaluated numerically by varying some or all single-particle parameters of the reference system. In Refs.~\onlinecite{koller_variational_2006, aichhorn_antiferromagnetic_2006, knap_jcl_2010} 
it was pointed out that the particle number of a certain particle species is thermodynamically consistent only if the corresponding chemical potential is considered as variational parameter. However, there is a formal difficulty when considering, for instance, the chemical potential $\mu$ or the energy spacing of the two-level systems $\epsilon$ as variational parameter since both couple to atomic excitations, which cannot be regarded as noninteracting particles. Generally a variation in these terms is not allowed within VCA, yet, this subtlety can be circumvented by mapping the atomic excitations onto hard-core bosons,
\[ S_i^+ \, \rightarrow \, b_i^\dagger \qquad S_i^- \, \rightarrow \, b_i^\nag \]
\[ \ket{-S}_i \, \rightarrow \, \ket{0}_i,\,\ket{-S+1}_i \, \rightarrow \, \ket{1}_i\,\ldots\,\ket{S}_i \, \rightarrow \, \ket{N}_i\;. \]
With this mapping the TCL Hamiltonian is given by
\begin{align}
 \hat{H}^{TCL} = &-t \sum_{\left\langle i,\,j \right\rangle} a_i^\dagger \, a_j^\nag -\Delta \sum_i b_i^\dagger \, b_i^\nag - (\mu-\omega_c)\,\hat{N}_p   \nonumber\\
 & +g\sum_i  ( a_i^\nag\,b_i^\dagger\,f_i^+  + a_i^\dagger\,b_i^\nag f_i^- ) \nonumber \\
 & + \lim_{U \rightarrow \infty} \frac{U}{2} \sum_i \prod_{\nu=0}^{N} (b_i^\dagger \, b_i^\nag - \nu)  \;\mbox{,}
 \label{eq:jclhc}
\end{align}
where $f_i^\pm=\sqrt{S(S+1)-(b_i^\dagger\,b_i^\nag - S) (b_i^\dagger\,b_i^\nag - S\pm1))}$. In \eq{eq:jclhc} we have formally added the hard-core constraint by introducing the infinite interaction $U$. Due to the fact that the Hamiltonian consists now of noninteracting particles the variation in the chemical potential $\mu$, which is needed to guarantee that the total particle number $\hat{N}_p$ is thermodynamically consistent, becomes possible.

In our calculations we varied the on-site energies and the hopping strength. In fact only two of the three on-site energies $\omega_c$, $\epsilon$, and $\mu$ are linear independent and therefore a combination of any  two of them used as variational parameters yields identical results. In particular, we use $\mat x = \lbrace \omega_c,\,\epsilon, \,t\rbrace$ as variational parameters, which ensures thermodynamic consistency for the particle number of both particle species and thus consistency for the total particle number as well.

The reference system defined on a cluster of size $L$ is solved using the band Lanczos method.\cite{freund_roland_band_2000, aichhorn_variational_2006}
The initial vector of the iterative band Lanczos method, which is used to evaluate the particle term of the Green's function, consists of $2L$ elements and is given by $\lbrace a_1^\dagger \ket{\psi_0},\,a_2^\dagger \ket{\psi_0} \,\ldots\, a_L^\dagger \ket{\psi_0},\,b_1^\dagger \ket{\psi_0},\,b_2^\dagger \ket{\psi_0}\,\ldots\,b_L^\dagger \ket{\psi_0} \rbrace$, where $\ket{\psi_0}$ is the $N_p$-particle ground state. In order to evaluate the hole term of the Green's function the creation operators are replaced by annihilation operators. Using the bosonic $\mat Q$-matrix formalism\cite{knap_spectral_2010}
we evaluate the grand potential and the single-particle Green's function of the physical system. As the TCL model consists of two distinct particle species, we extract a Green's function for photons $G^{ph}(\ve k,\,\omega) \equiv G_{a_\ve{k}^\nag a_\ve{k}^\dagger}(\omega)$ and a Green's function for the atomic excitations $G^{ex}(\ve k,\,\omega) \equiv G_{b_\ve{k}^\nag b_\ve{k}^\dagger}(\omega)$. From the Green's functions we evaluate the single-particle spectral function
\begin{equation}
 A^x(\ve{k},\,\omega)\equiv-\frac{1}{\pi} \mbox{Im} \, G^x(\ve{k},\,\omega)\;\mbox{,}
 \label{eq:spe:spectralfunction}
\end{equation}
the density of states
\begin{equation}
 N^x(\omega)\equiv \int A^x(\ve{k},\,\omega) \, d\ve{k}  = \frac{1}{N} \sum_{\ve{k}} A^x(\ve{k},\,\omega)
 \label{eq:spe:dos}
\end{equation}
and the momentum distribution
\begin{equation}
 n^x(\ve{k})\equiv - \int_{-\infty}^0 A^x(\ve{k},\,\omega)\,d\omega \;\mbox{,}
\end{equation}
where $x$ can be either $ph$ for photons or $ex$ for atomic excitations.

\section{\label{sec:qpt}Quantum phase transition}
The nonlinearities which arise due to the coupling of the photons to the ensemble of two-level systems lead to a quantum phase transition from Mott phase to superfluid phase. The elementary excitations in light-matter systems---the polaritons---are linear combinations of photons and atomic excitations. The Mott phase is characterized by integer polariton density, zero compressibility, and a gap in the spectral function. 
Intriguingly, polaritons in Mott phase are localized in cavities, which in turn means that the photons are not able to tunnel to adjacent cavities, since too much energy would be needed for this process. Hence, the Mott phase can be considered as a stable state of frozen light. In superfluid phase, however, the polaritons are delocalized on the whole lattice and Bose condense in the state of zero momentum. 

\begin{figure}
	\centering
	\includegraphics[width=0.48\textwidth]{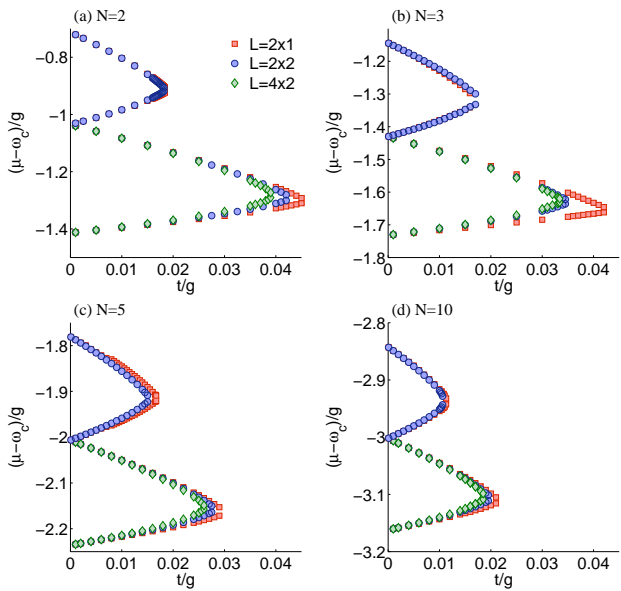}
	\caption{(Color online) Phase boundary of the two-dimensional TCL model evaluated for zero detuning $\Delta=0$, reference systems of size $L$, and $N$ two-level systems per cavity, where in \fc{a} $N=2$, \fc{b} $N=3$, \fc{c} $N=5$, and \fc{d} $N=10$. }
	\label{fig:pd}
\end{figure}

\begin{figure}
	\centering
	\includegraphics[width=0.3\textwidth]{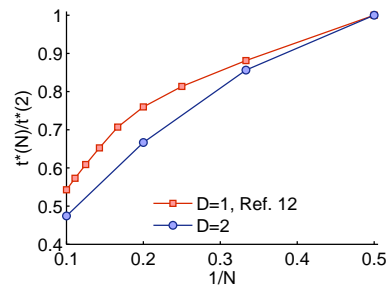}
	\caption{(Color online) Critical hopping strength ratio $t^*(N)/t^*(2)$ for the first Mott lobe in dependence on the dimension $D$. Results for one dimension are obtained from Ref.~\onlinecite{rossini_mott-insulating_2007}. }
	\label{fig:comp}
\end{figure}

We determine the phase boundary of the two-dimensional TCL model for zero detuning $\Delta=0$. The first two Mott lobes with polariton density $n_p=1$ and $n_p=2$, respectively, are shown in \fig{fig:pd} for distinct values of the two-level system number $N=\lbrace 2,\,3,\,5,\,10\rbrace$.
The boundary of the quantum phase transition is given by the minimal amount of energy necessary to add (remove) a particle to (from) the system and can therefore be evaluated directly from the minimal gap of the single-particle spectral function obtained by means of VCA. 
The size of the gap does not depend on the particle species the spectral function is evaluate for, since photons and atomic excitations are coupled by $g$. 
As already mentioned in \Se.~\ref{sec:method}, we use the variational parameters $\mat x = \lbrace \omega_c,\,\epsilon, \,t\rbrace$, which allow for thermodynamic consistency in the total polariton number. In contrast to the results in one dimension\cite{rossini_mott-insulating_2007}
the lobes are round shaped and no reentrance behavior can be observed for increasing hopping strength $t$. For increasing number of two-level systems $N$ the width of the Mott lobes with different polariton density $n_p$ becomes more similar, see \tab{tab:ratio}.
\begin{table}
        \caption{ Ratio $w_1/w_i$ of the width of Mott lobe $1$ with polariton density $n_p=1$ and Mott lobe $i$ with polariton density $n_p=i$ for $N$ two-level systems per cavity. }
        \label{tab:ratio}
        \centering
        \begin{tabular}{cccc}
                \hline
                \hline
                {$\quad N\quad $} & $\quad w_1/w_2\quad $ & $\quad w_1/w_3\quad $ &$\quad w_1/w_4\quad $\\
                \hline
		2 & 1.18 & 2.84 & 4.83 \\
		3 & 1.05 &   1.38  &  2.26 \\
		5 & 1.02  &  1.09   & 1.25 \\
		10 & 1.00  &  1.02 &   1.04  \\
                \hline
                \hline
        \end{tabular}
\end{table}
The critical hopping strength $t^*$, which determines the tip of the Mott lobes, depends on both the filling $n_p$ and the number of two-level systems per cavity $N$. In particular, $t^*(N)$ is shrinking for increasing $N$ for a fixed polariton density $n_p$. In \fig{fig:comp} we investigate for the first Mott lobe, \ie, for $n_p=1$, the dimensionality dependence of the ratio $t^*(N)/t^*(2)$, which specifies how fast the lobes are shrinking with increasing $N$. 
To this end, we compare our VCA results for two dimensions with DMRG results for one dimension obtained by D.~Rossini \textit{et al.} in Ref.~\onlinecite{rossini_mott-insulating_2007}. It can be observed that with increasing $N$ the lobes are shrinking faster in two dimensions than in one dimension. 

The phase boundary at zero hopping can be determined analytically, as the model decouples into $M$ single-cavity problems, \ie, into $M$ TC systems shifted by the chemical potential $-\mu \hat{n}_p$. The TC model has been solved exactly for zero detuning\cite{Tavis_Exact_Solution1968,Tavis_Approximate_Solutions_1969}
and for nonzero detuning.\cite{Tavis_Approximate_Solutions_1969,Bogoliubov_exact_1996}
Since the full analytic solution is involved we concentrate here in determining the zero hopping phase boundary of the first Mott lobe for zero detuning, which is relevant for our data. To this end, we diagonalize the TC model for $n_p=0,\,1,\,\text{and}\,2$ polaritons which yields
\begin{subequations}
\begin{align}
 E_{0} &= 0  \\
 E_{1} &= \lbrace (\omega_c - \mu) \pm \sqrt{N}\rbrace \\
 E_{2} &= \lbrace 2(\omega_c - \mu) \pm \sqrt{2(2N-1)},\,2(\omega_c - \mu)\rbrace\;\text{.} 
\end{align}
\label{eq:entc}
\end{subequations}
The number of eigenstates for the sector of $n_p$ polaritons is $n_p+1$ if $n_p<N$ and $N+1$ if $n_p \geq N$.\cite{na_strongly_2008, yamamoto_mesoscopic_1999}
The phase boundary is evaluated by comparing the ground-state energies of adjacent sectors, which yields $\mu - \omega_c = - \sqrt{N}$ for the boundary between $n_p=0$ and $n_p=1$ and $\mu - \omega_c = \sqrt{N} -\sqrt{2(2N-1)} $ for the boundary between $n_p=1$ and $n_p=2$. This is of course in full agreement with our numerical results. In light-matter systems the optical nonlinearities arise as the energy which is needed to add the first excitation to the system is smaller than the one to add the second excitation. This results in a repulsive interaction of size $2\sqrt{N} - \sqrt{2(2N-1)}$, which is approximately $1/2\sqrt{N}$ for large $N$.

\section{\label{sec:excitations}Excitations}
In this section we investigate the excitations of the TCL model. In particular, we evaluate single-particle spectral functions and densities of states of photons and atomic excitations. Furthermore we present the momentum distribution for both particle species. Based on the spectral information we characterize polaritons, which are the elementary excitations in light-matter systems.

\subsection{Spectral properties of photons and atomic excitations}
\begin{figure}
	\centering
	\includegraphics[width=0.48\textwidth]{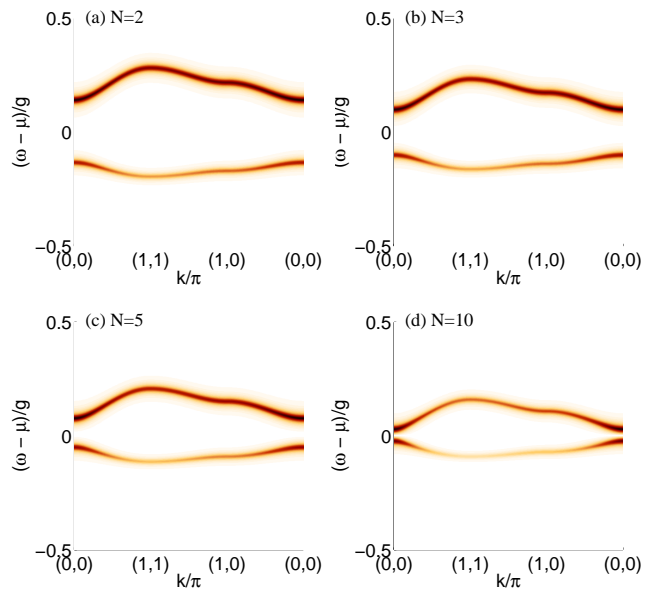}
	\caption{(Color online) Hole band $\omega_h$ and the lowest-lying particle band $\omega_p^1$ of the photon single-particle spectral function $A^{ph}(\ve k,\,\omega)$ for fixed hopping strength $t=0.015$ and zero detuning $\Delta=0$. The modified chemical potential and the number of two-level systems is \fc{a} $\mu-\omega_c=-1.25$, $N=2$, \fc{b} $\mu-\omega_c=-1.6$, $N=3$, \fc{c} $\mu-\omega_c=-2.15$, $N=5$, and \fc{d} $\mu-\omega_c=-3.1$, $N=10$.  }
	\label{fig:sf}
\end{figure}

Photon spectral functions $A^{ph}(\ve k,\,\omega)$ evaluated by means of VCA for fixed hopping strength $t=0.015$ and zero detuning $\Delta=0$ are shown in \fig{fig:sf} for $N=\lbrace 2,\,3,\,5,\,10 \rbrace$ two-level systems located in each cavity. For increasing number of two-level systems $N$ the Mott lobes shrink and thus the gap in the spectral function is decreasing. 
The modified chemical potential $\mu-\omega_c$ is chosen such that the spectral function is evaluated approximately in the middle of the Mott lobe. We used the variational parameters $\mat{x}=\lbrace \omega_c,\,\epsilon,\,t \rbrace$, reference systems of size $L=4\times 2$, and an artificial broadening $0^+=0.01$ for the numerical evaluation.

The number of particle and hole bands, present in the single-particle spectral function and their approximate energies can be already determined from the single-cavity solution. For large enough filling ($n_p>N$) there are $N+1$ eigenstates in the sectors of $n_p\pm1$ particles which leads to $N+1$ particle and hole bands, respectively. However, we investigate spectral properties in the first Mott lobe ($n_p=1$) and thus we have to examine the zero- and two-particle sectors. The zero-particle sector ($n_p=0$) consists of only one state leading to one hole band $\omega_h$ and the two-particle sector ($n_p=2$) consists for all $N\geq 2$ of three states, which leads to three particle bands $\omega_p^i$, where $i \in \lbrace 1,\,2,\,3 \rbrace$. We choose the order of the bands such that the excitation energy increases with increasing index $i$. For clarity \fig{fig:sf} shows only the hole band $\omega_h$ and the lowest-lying particle band $\omega_p^1$.
The approximate location of the particle bands is obtained from the energy difference of the eigenenergies of the two particle sector and the ground-state energy of the one-particle sector, see \eq{eq:entc}, leading to
{\allowdisplaybreaks
\begin{align*}
 \omega_p^1 &\approx (\omega_c - \mu) - \sqrt{2(2N-1)}+\sqrt{N} \\
 \omega_p^2 &\approx (\omega_c - \mu) + \sqrt{N} \\
 \omega_p^3 &\approx (\omega_c - \mu) + \sqrt{2(2N-1)}+\sqrt{N} \;\text{.}
\end{align*}}
Analogously, one obtains for the hole band
\begin{equation*}
 \omega_h \approx (\omega_c - \mu) - \sqrt{N} \;\text{.}
\end{equation*}

\begin{figure}
	\centering
	\includegraphics[width=0.48\textwidth]{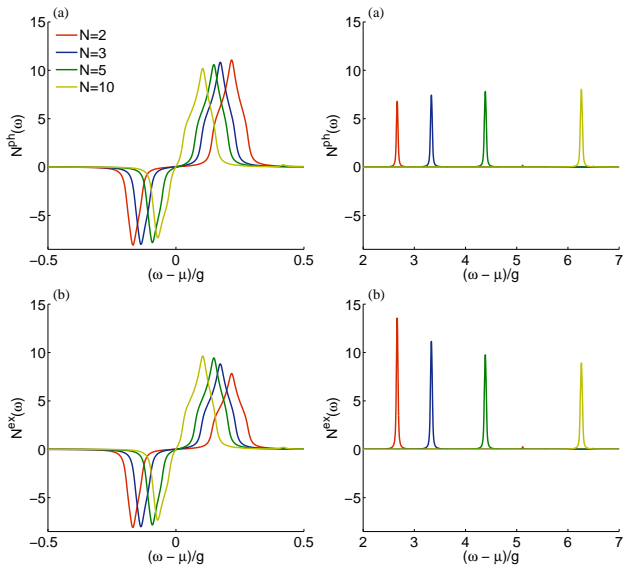}
	\caption{(Color online) Density of states of \fc{a} photons $N^{ph}(\omega)$ and \fc{b} atomic excitations $N^{ex}(\omega)$ for parameters as in \fig{fig:sf}. The left column shows contributions from the bands $\omega_h$ and $\omega_p^1$ and the right column from the bands with higher energy $\omega_p^2$ and some of the $\omega_p^3$. }
	\label{fig:dos}
\end{figure}

The densities of states of both photons $N^{ph}(\omega)$ as well as atomic excitations $N^{ex}(\omega)$ are shown in the first and second row, respectively, of \fig{fig:dos} for identical parameters as in the case of the single-particle spectral function. 
The left column contains the density of states of a small energy window centered around zero, showing the low-lying excitation bands $\omega_h$ and $\omega_p^1$. The right column contains data for higher excitation energies. The bands $\omega_p^2$ carry significant spectral weight whereas the bands $\omega_p^3$ are barely visible for $N=2$ ($\omega_p^3 \approx 5.1$) and $N=3$ ($\omega_p^3\approx6.5$). For more than three two-level systems per cavity the excitation energy is already larger than the maximum energy considered in the plot. The position of the bands matches well with the approximate results obtained from the single-cavity limit. In the photon density of states $N^{ph}(\omega)$, first row, the spectral weight of the low-lying particle band $\omega_p^1$ decreases with increasing two-level system number $N$, whereas the spectral weight of $\omega_p^2$ increases. The opposite is true for the atomic excitation density of states $N^{ex}(\omega)$, second row. Interestingly, due to this behavior photon and atomic excitation densities of states become more similar for increasing two-level system number $N$, which is due to the fact that the ensemble of two-level systems behaves more like free bosons for large $N$.\cite{na_strongly_2008, yamamoto_mesoscopic_1999}

\begin{figure}
	\centering
	\includegraphics[width=0.48\textwidth]{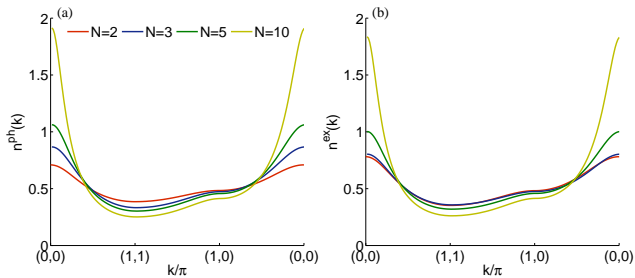}
	\caption{(Color online) Momentum distribution of \fc{a} photons $n^{ph}(\ve k)$ and \fc{b} atomic excitations $n^{ex}(\ve k)$ for parameter as in \fig{fig:sf}. }
	\label{fig:n}
\end{figure}

The momentum distribution of both particle species, which can be evaluated with high accuracy by means of the $\mat Q$-matrix formalism,\cite{knap_spectral_2010} 
is shown in \fig{fig:n}.
The momentum distribution of photons (left panel) and the atomic excitations (right panel) do not exhibit major differences. For increasing number of two-level systems $N$ and fixed hopping strength $t$, the tip of the Mott lobe is approached, see \fig{fig:pd}. 
From this in turn it follows that the density in the center of the Brillouin zone is increasing for increasing $N$, which is in accordance with our results.

\subsection{Polaritons}
Here, we investigate properties of polaritons, the elementary excitations of the TCL model, which are linear combinations of photons and atomic excitations. Our goal is to describe these excitations by polaritonic quasiparticles added to the $N_p$-particle ground state $\ket{\psi_0}$. Hence, we introduce the polariton creation operators $p_{\alpha,\ve k}^\dagger$ for particle excitations and $h_{\alpha,\ve k}^\dagger$ for hole excitations as suitable linear combinations of photons and hard-core bosons introduced in \se~\ref{sec:method},
\begin{subequations}
\begin{align}
 p_{\alpha,\ve{k}}^\dagger &= \beta_p^\alpha(\ve{k})\,a_\ve{k}^\dagger + \gamma_p^\alpha(\ve{k})\,b^\dagger_\ve{k} \;\mbox{,} \\
 h_{\alpha,\ve{k}}^\dagger &= \beta_h^\alpha(\ve{k})\,a^\nag_\ve{k} + \gamma_h^\alpha(\ve{k})\,b^\nag_\ve{k} \;\mbox{.}
\end{align}
\label{eq:polCreationOps}
\end{subequations}
The weights $\beta_p^\alpha(\ve{k})$ and $\gamma_p^\alpha(\ve{k})$ of the polariton creation operators depend on the wave vector $\ve k$, the quasiparticle band index $\alpha$, and the filling $n_p$. The dependence on the latter is not explicitly included in the notation as we solely focus on the first Mott lobe with particle density $n_p=1$. It is important to notice that the hole creation operator is neither the adjoint of the particle creation operator nor its annihilation counterpart. The normalized polariton quasiparticle states are generated by applying the polariton particle and hole creation operators on the $N_p$-particle ground state,
{\allowdisplaybreaks
\begin{subequations}
\begin{align}
 \ket{\tilde{\psi}_{p,\ve{k}}^\alpha} &= \frac{p_{\alpha,\ve{k}}^\dagger \ket{\psi_0} }{ \sqrt{\bra{\psi_0} p_{\alpha,\ve{k}} \, p_{\alpha,\ve{k}}^\dagger \ket{\psi_0} }} \;\mbox{,} \\
 \ket{\tilde{\psi}_{h,\ve{k}}^\alpha} &= \frac{h_{\alpha,\ve{k}}^\dagger \ket{\psi_0} }{ \sqrt{\bra{\psi_0} h_{\alpha,\ve{k}} \, h^\dagger_{\alpha,\ve{k}} \ket{\psi_0}}} \;\mbox{.}
\end{align}
\label{eq:polwavefu}
\end{subequations}}
The weights $\beta$ and $\gamma$ of the linear combination are determined by maximizing the overlap between the exact eigenvectors $\ket{\psi^{N_p \pm 1}_{\alpha,\ve k}}$ of the TCL model in the $(N_p\pm1)$-particle sector. This yields a generalized eigenvalue problem which is used to determine the weights $\beta$ and $\gamma$, see Ref.~\onlinecite{knap_jcl_2010}
for a detailed derivation and discussion. The eigenvalue $\lambda$ of the generalized eigenvalue problem specifies the quality of the quasiparticle description. More specifically, $\lambda$ is bound by the interval $[0,\,1]$, where $\lambda=1$ corresponds to a perfect description by polariton quasiparticles, \ie, to maximal overlap between the true $(N_p \pm 1)$-particle states and the polariton states $\ket{\tilde{\psi}_{p/h,\ve{k}}^\alpha}$, whereas small values of $\lambda$ indicate a modest quasiparticle description. The generalized eigenvalue problem fixes the weights $\beta$ and $\gamma$ only upon a constant, which is determined by the condition that the total spectral weight consisting of the spectral weight of photons and atomic excitations is conserved.

\begin{figure}
	\centering
	\includegraphics[width=0.4\textwidth]{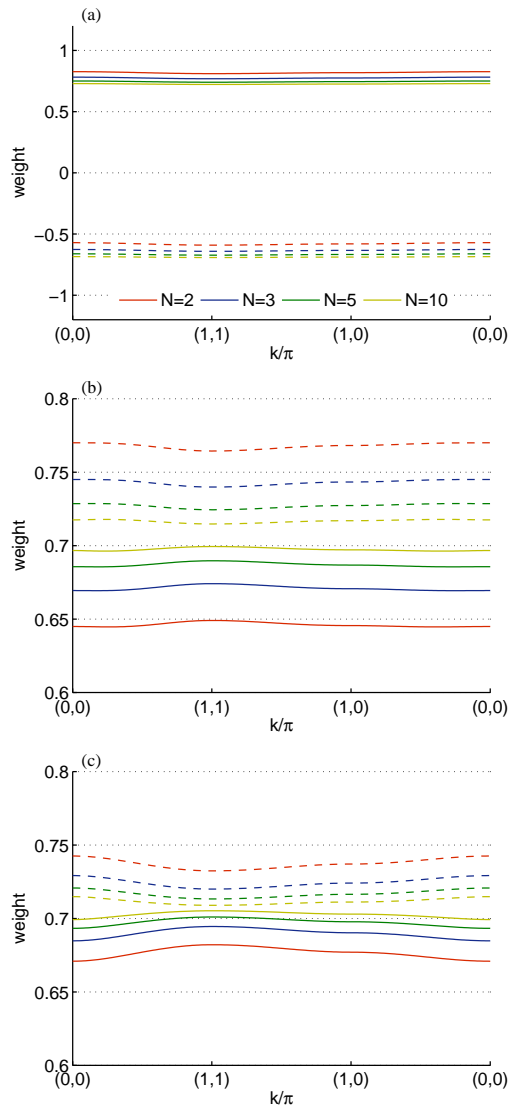}
	\caption{(Color online) Photon weights $\beta$ (solid lines) and atomic excitation weights $\gamma$ (dashed lines) of the polaritonic quasiparticle creation operators $p_{\alpha,\ve{k}}^\dagger$ for the bands \fc{a} $\omega_p^1$, \fc{b} $\omega_p^2$, \fc{c} $\omega_p^3$, and $N=\lbrace 2,\,3,\,5,\,10 \rbrace$ two-level systems per cavity. }
	\label{fig:qp}
\end{figure}

For the hole band $\omega_h$ the weights $\beta$ and $\gamma$ can be chosen freely, as both $a_{\ve k}$ and $b_{\ve k}$ applied on the ground state with particle density $n_p=1$ are proportional to the same state. Thus we investigate only the weights $\beta$ and $\gamma$ for the particle bands $\omega_p^1$, $\omega_p^2$, and $\omega_p^3$, which are shown in \fig{fig:qp} from top to bottom for $N=\lbrace 2,\,3,\,5,\,10 \rbrace$ two-level systems per cavity. 
The photon weight $\beta$ corresponds to the solid line and the atomic-excitation weight to the dashed line. For the band with lowest excitation energy $\omega_p^1$ the weights $\beta$ and $\gamma$ are of opposite sign, whereas the sign is equal for the bands $\omega_p^2$ and $\omega_p^3$. The bands $\omega_p^1$ and $\omega_p^2$ are very well described by the polariton picture as $\lambda \approx 1$. Yet, the band with highest energy $\omega_p^3$ is very poorly represented by the polariton creation operators as $\lambda \approx 0.01$. For increasing number of two-level systems $N$ per cavity the weights of the two constituents become similar. This might indicate, as in the case of the spectral weight, that the atomic excitations behave for large $N$ like bosonic particles. In the case of a single two-level system per cavity $N=1$, \ie, the Jaynes-Cummings lattice model, the asymmetry in the coefficients is much more pronounced than it is here.\cite{knap_spectral_2010}

\section{\label{sec:conclusion}Conclusions}
We presented and discussed the quantum phase transition and the excitations of the Tavis-Cummings lattice model in two dimensions obtained within the variational cluster approach. The Tavis-Cummings lattice model describes light-matter systems which contain multiple two-level structures in each cavity. Due to this fact the Tavis-Cummings lattice model might be easier to realize in the experiment than its counterpart, the Jaynes-Cummings lattice model, which contains exactly one two-level systems per cavity. As a goal for future research, a detailed study of cavities with a random number of two-level systems might provide further interesting insight into light-matter systems. 

In this paper, we determined the quantum phase transition delimiting Mott phase, in which polaritons are localized in each cavity, from superfluid phase, in which polaritons are delocalized on the whole lattice. We studied the dependence of this phase boundary for various two-level system numbers per cavity. For increasing number of two-level systems the Mott lobes become narrower, however, the width of  the Mott lobes for distinct filling becomes more equal. We also compared the dependence of the critical hopping strength, which determines the tip of the Mott lobe, on the dimension of the coupled cavity system. Additionally to the phase boundary, we investigated spectral functions and corresponding densities of states. The variational cluster approach allows us to extract spectral properties for both photons as well as atomic excitations, provided the latter are mapped onto hard-core bosons. By investigating the zero-hopping limit, which corresponds to investigating a single cavity, we determine the number of bands in the spectral function and their approximate location. For the first Mott lobe there exist three particle bands and one hole band. The particle band and the hole band with smallest energy are reminiscent of the excitations in the Bose-Hubbard model. Particularly, they are also cosine shaped and the density distribution of the weight is similar.\cite{elstner_dynamics_1999, sengupta_mott-insulator-to-superfluid_2005, huber_dynamical_2007, knap_spectral_2010, pippan_2010} The additional two particle bands lie at considerably higher energies. The band with second highest excitation energy carries significant spectral weight whereas the one with highest energy is barely visible in the spectra. Interestingly, for increasing two-level system number the weight of the photon spectra becomes more and more similar to the weight of the atomic excitation spectra. We investigated the momentum distribution as well, which is rather similar for photons and atomic excitations. Yet, for increasing two-level system number and constant hopping strength, a larger density can be observed in the center of the Brillouin zone, which arises due to the fact that the Mott lobe is shrinking with increasing number of two-level systems. Therefore, for identical hopping strength the boundary to superfluid phase is approaching, which is responsible for the increasing density in the center of the Brillouin zone. Finally, we studied the properties of polaritons, the elementary excitations in light-matter systems. Since we evaluated spectral properties of both particle species, we were able to introduce polariton quasiparticle and quasihole creation operators as linear combinations of photons and atomic excitations. The polariton creation operators depend on the wave vector, band index and filling. We investigated the photon and atomic excitation weights of the linear combination and analyzed their dependence on the number of two-level systems located in each cavity.

\begin{acknowledgments}
We are grateful to D.~Rossini for providing us the data used in \fig{fig:comp}. We made use of parts of the ALPS library (Ref.~\onlinecite{albuquerque_alps_2007}) for the implementation of lattice geometries and for parameter parsing.
We acknowledge financial support from the Austrian Science Fund (FWF) under the doctoral program ``Numerical Simulations in Technical Sciences'' Grant No. W1208-N18 (M.K.) and under Project No. P18551-N16 (E.A.).
\end{acknowledgments}

\end{document}